\documentclass[dvips]{article}
\usepackage{icrctc07}

\title{The Baikal Neutrino Telescope: Selected Physics Results}
\shorttitle{Short title}

\authors{K. Antipin$^{1}$, V. Aynutdinov$^{1}$, V. Balkanov$^{1}$, 
I. Belolaptikov$^{4}$, N. Budnev$^{2}$, I. Danilchenko$^{1}$, G. Domogatsky$^{1}$,
A. Doroshenko$^{1}$, A. Dyachok$^{2}$, Zh. Dzhilkibaev$^{1}$, S. Fialkovsky$^{6}$,
O. Gaponenko$^{1}$, K. Golubkov$^{4}$, O. Gress$^{2}$, T. Gress$^{2}$, O. Grishin$^{2}$, A. Klabukov$^{1}$,
A. Klimov$^{8}$, A. Kochanov$^{2}$, K. Konischev$^{4}$, A. Koshechkin$^{1}$, V. Kulepov$^{6}$, L. Kuzmichev$^{3}$,
E. Middell$^{5}$, S. Mikheyev$^{1}$, M. Milenin$^{6}$, R. Mirgazov$^{2}$, E. Osipova$^{3}$, Yu. Pavlova$^{1}$,
G. Pan'kov$^{2}$,
L. Pan'kov$^{2}$, A. Panfilov$^{1}$, D. Petukhov$^{1}$, E. Pliskovsky$^{4}$, P. Pokhil$^{1}$, V. Poleshuk$^{1}$,
E. Popova$^{3}$, V. Prosin$^{3}$, M. Rosanov$^{7}$, V. Rubtzov$^{2}$, B. Shaibonov$^{4}$, A. Sheifler$^{1}$,
A. Shirokov$^{3}$, Ch. Spiering$^{5}$, B. Tarashansky$^{2}$, R. Wischnewski$^{5}$, I. Yashin$^{3}$,
V. Zhukov$^{1}$
}
\shortauthors{Author and et al.}
\afiliations{$^1$Institute for Nuclear Research, Moscow, Russia\\ 
$^2$Irkutsk State University, Irkutsk, Russia\\
$^3$Skobeltsyn Institute of Nuclear Physics  MSU, Moscow, Russia\\
$^4$Joint Institute for Nuclear Research, Dubna, Russia\\
$^5$DESY, Zeuthen, Germany\\ 
$^6$Nizhni Novgorod State Technical University, Nizhni Novgorod, Russia\\ 
$^7$St Petersburg State Marine University, St Petersburg, Russia\\ 
$^8$Kurchatov Institute, Moscow, Russia\\ 
}
\email{wischnew@ifh.de}

\abstract{We present results on searches for exotic particles
(relativistic magnetic monopoles and WIMPs) and for UHE neutrinos, 
obtained with the Baikal
neutrino telescope NT200.
}

\begin{document}
\maketitle
\section{Introduction}

The Baikal Neutrino Telescope  is operated in Lake 
Baikal, Siberia,  at a depth of \mbox{1.1 km}. 
The first stage telescope configuration  NT200 
was put into permanent operation 
on April 6th, 1998.
The upgraded telescope NT200+ was put into operation on April 9th, 2005. 
This configuration consists of the old NT200 telescope,
surrounded by three new external strings.
Status and
description of the detector has been presented elsewhere ~\cite{APP2}. 
In this paper we present selected physics results of a search for 
fast magnetic monopoles, neutrino signals from WIMP annihilation at the
center of the Earth and a search for diffuse neutrinos with energies 
larger than 10 TeV, obtained with the Baikal neutrino telescope NT200.

\section{Fast Magnetic Monopoles}
Fast magnetic monopoles with Dirac charge $g=68.5 e$ are 
interesting objects to search for with deep underwater neutrino telescopes.
The intensity of monopole Cherenkov radiation is $\approx$ 8300
times higher than that of muons. Optical modules of the 
Baikal experiment can detect such an object from a distance up 
to hundred meters.  
The processing chain for fast monopoles starts with the selection of
events with a high multiplicity of hit channels: $N_{hit}>30$. 
In order to reduce  the background  from downward atmospheric muons 
we restrict ourself to monopoles coming from the lower hemisphere. 
For an upward going particle the times of hit channels increase 
with rising $z$-coordinates from bottom to top of the detector. 
To suppress events caused by downward  moving particles, a cut on the value of the 
time--$z$--correlation, $C_{tz}$, is applied:             
\begin{equation}
  \label{eq1}
  C_{tz}= \frac{ \sum_{i=1}^{N_{hit}}(t_i- \overline t )(z_i- \overline
    z)} {N_{hit} \sigma_t \sigma_z}>0
\end{equation}
where $t_i$ and $z_i$ are time and $z$-coordinate of a fired channel, 
$\overline t$ and $\overline z$ are mean values for times and
$z$-coordinates of the event and $ \sigma_t$ and $ \sigma_z$ the 
rms--errors for time and $z$-coordinates.
\begin{figure}
\begin{center}
\noindent
\includegraphics [width=0.4\textwidth]{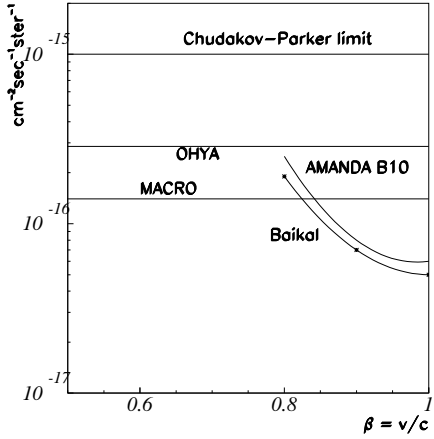}
\end{center}
\caption{Upper limits on the flux of fast monopoles
obtained in this analysis (Baikal) and in other 
experiments.}\label{fig1}
\end{figure}


Within 1003 days of live time used in this analysis, about 
$ 3 \cdot 10^8$ events with $N_{hit} > 4 $ have been recorded, 
with 21240 of them satisfying conditions $N_{hit}>30$ and $C_{tz}>0$. 
For further background suppression 
(see ~\cite{MONO06} for details of the analysis)
we use additional cuts, which essentially reject muon events 
and at the same time only slightly reduce
the effective area for relativistic monopoles.

No events from the experimental sample pass all cuts.
For the time periods included in the analysis the acceptances $ A_{eff} $ varies between
$3 \cdot 10^8$ and $6 \cdot 10^8 $ cm$^2$sr (for $ \beta=1$).
From the non-observation of candidate events in NT200 and  the earlier
stage telescopes NT36 and NT96 \cite{baikal},  
a combined $90\%$ C.L. upper limit on the flux of fast monopoles 
is obtained. In Fig.~\ref{fig1} we compare this upper limit 
to the limits from the experiments Ohya,
MACRO and AMANDA \cite{ohya}.        
The Baikal limit is currently the most stringent one.
\begin{figure}
\begin{center}
\noindent
\includegraphics [width=0.45\textwidth]{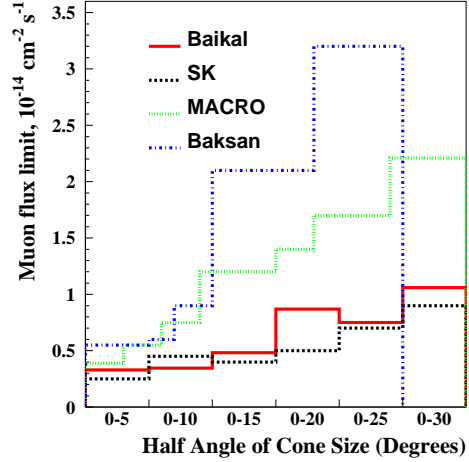}
\end{center}
\caption{Limits on the excess muon flux from the center of 
the Earth versus half-cone of the search angle.}\label{fig2}
\end{figure}

\section{WIMPs}
The search for WIMPs with the Baikal
neutrino telescope is based on a possible signal of
nearly vertically upward going muons, exceeding
the flux of atmospheric neutrinos. 
The method of event selection relies on 
the application of a series of cuts which are tailored to the response
of the telescope to nearly vertically upward moving muons. 

For the present analysis
we included all events with $\ge$5 hit channels, out of which 
$\ge$4 hits are along at least one of all hit strings.
To this sample, a series of 5 cuts is applied
(see \cite{BWIMP} for details of analysis). 
The applied cuts select muons with -1$< \cos(\theta) <$-0.65 
and result in a detection area of about 1800 m$^2$ for vertically 
upward going muons with energies $>10$ GeV.

From 1038 days of effective data taking 
between April 1998 and February 2003,
48 events with -1$< \cos(\theta) <$-0.75 have been selected 
as neutrino candidates, compared to 56.6 events expected from atmospheric neutrinos
in case of oscillations and 73.1 without oscillations. 
Within statistical uncertainties the experimental
angular distribution is  consistent with the prediction
including neutrino oscillations.

\begin{figure}
\begin{center}
\noindent
\includegraphics [width=0.45\textwidth]{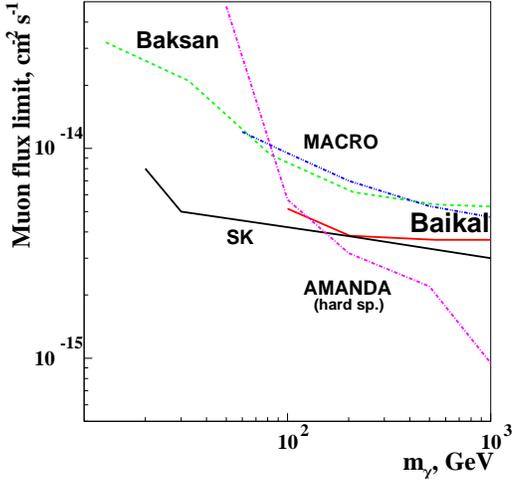}
\end{center}
\caption{Limits on the excess muon flux from the center of 
the Earth as a function of WIMP mass.}\label{fig3}
\end{figure}

Regarding the 48 detected events as being induced by atmospheric 
neutrinos, one can derive an upper limit on the additional flux of muons 
from the center of the Earth due to annihilation of neutralinos - the 
favored candidate for cold dark matter. The 90\% C.L. muon flux limits 
for six cones around the opposite zenith obtained with NT200 
($E_{\mbox{thr}}>$10 GeV) in 1998-2002 are shown in Fig.~\ref{fig2}.
It was shown \cite{BAKSANWIMP} that the size of a cone 
which contains 90\% of signal strongly depends on neutralino mass.
The 90\% C.L. flux limits are calculated as a function of neutralino 
mass using cones which collect 90\% of the expected signal 
and  are corrected for the 90\% collection efficiency due to cone size.
Also a correction is applied for each neutralino mass to translate
from the threshold of the 10 GeV to the 1 GeV threshold.
These limits are shown in Fig.~\ref{fig3}. Also shown in Fig.~\ref{fig3}
are limits obtained by 
Baksan, MACRO, 
Super-Kamiokande and AMANDA \cite{BAKSANWIMP}.

\section{UHE neutrinos}
The BAIKAL survey for high energy neutrinos searches
for bright cascades produced at the neutrino interaction
vertex in a large volume around the neutrino telescope \cite{HEB}.
We select events with high multiplicity of hit channels $N_{\mbox{\small hit}}$,
corresponding to bright cascades. 
To separate high-energy neutrino events
from background events, a cut
to select events with upward moving light signals has been developed.
We define for each event
$t_{\mbox{\footnotesize min}}=\mbox{min}(t_i-t_j)$,
where $t_i, \, t_j$ are the arrival times at channels $i,j$ 
on each string, and the minimum over all strings is calculated.
Positive and negative values of $t_{\mbox{\footnotesize min}}$ correspond to 
upward and downward propagation of light, respectively. 

Within the 1038 days of the detector live time
between April 1998 and February 2003, 
$3.45 \times 10^8$ events with $N_{\mbox{\small hit}} \ge 4$ have been recorded. 
For this analysis we used 22597 events with hit channel multiplicity
$N_{\mbox{\small hit}}>$15 and  
$t_{\mbox{\footnotesize min}}>$-10 ns.
We conclude that 
data are consistent with simulated background
for both $t_{\mbox{\footnotesize min}}$ and $N_{\mbox{\small hit}}$ 
distributions. No statistically significant excess above the background 
from atmospheric muons has been observed. 

\begin{table}[htb]
\label{tab1}
\begin{center}
\begin{tabular}{l|cc} \hline 
 & BAIKAL & AMANDA \\
\hline 
Model & $n_{90\%}/N_{\mbox{\footnotesize m}}$ & 
$n_{90\%}/N_{\mbox{\footnotesize m}}$  \\
\hline
  10$^{-6}\times E^{-2}$ & 0.81 & 0.22  \\
  SS Quasar  & 0.25 & 0.21  \\
  SS05 Quasar  & 2.5 & 1.6  \\
  SP u  & 0.062 & 0.054  \\
  SP l & 0.37 & 0.28  \\
  P $p\gamma$ & 1.14 & 1.99  \\
  M $pp+p\gamma$ & 2.86 & 1.19  \\
  MPR & 4.0 & 2.0  \\
  SeSi  & 2.12 & -  \\
  \hline 
\end{tabular} 
\caption{
Model rejection factors for models of astrophysical neutrino sources.
}
\end{center}
\end{table}
\vspace{-0.5cm}
Since no events have been observed which pass the final cuts,
upper limits on the diffuse flux of extraterrestrial 
neutrinos are calculated. For a 90\% C.L. an upper limit 
on the number of signal events of $n_{90\%}=$2.5  is obtained 
assuming an uncertainty in signal detection of 24\% 
and a background of zero events.
A model of astrophysical neutrino sources, for which the total number
of expected events $(\nu_e+\nu_{\mu}+\nu_{\tau})$, $N_{\mbox{\footnotesize m}}$, is larger than 
$n_{90\%}$, is ruled out at 90\% C.L.. 
Table 1 represents model rejection factors (MRF) 
$n_{90\%}/N_{\mbox{\footnotesize m}}$ 
for models of astrophysical neutrino sources (see \cite{HEB} for references) 
obtained from our search, compared to AMANDA \cite{AMANDAHE}.

For an $E^{-2}$ behaviour of the neutrino spectrum and a flavor ratio 
$\nu_e:\nu_{\mu}:\nu_{\tau}=1:1:1$ at the Earth, the 90\% C.L. upper 
limit on the neutrino flux of all flavors obtained with the Baikal 
neutrino telescope  NT200 (1038 days) is \cite{HEB}:
\begin{equation}
E^2\Phi<8.1 \times 10^{-7} 
\mbox{cm}^{-2}\mbox{s}^{-1}\mbox{sr}^{-1}\mbox{GeV}.
\label{eq2}
\end{equation}
For the resonant process 
with the resonant neutrino energy  
$E_0=6.3\times 10^6 \,$GeV, 
the model-independent limit on $\bar{\nu_e}$ is \cite{HEB}: 
\begin{equation}
\Phi_{\bar{\nu_e}} < 3.3 \times 10^{-20}
\mbox{cm}^{-2}\mbox{s}^{-1}\mbox{sr}^{-1}\mbox{GeV}^{-1}.
\label{eq3}
\end{equation}

\begin{figure}
\begin{center}
\noindent
\includegraphics [width=0.45\textwidth]{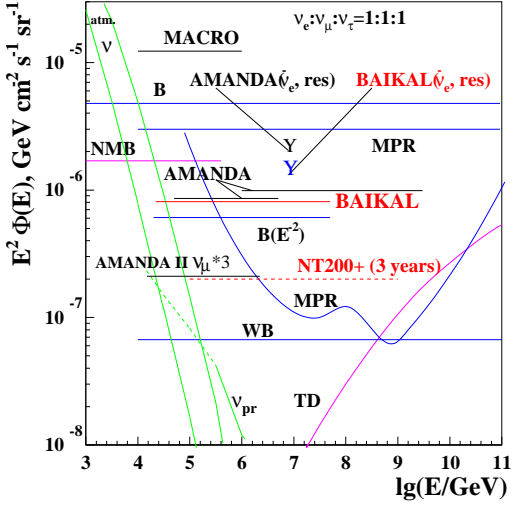}
\end{center}
\caption{All-flavor neutrino flux 
predictions in different models of neutrino sources 
compared to experimental upper limits to $E^{-2}$ fluxes obtained
by this analysis and other experiments (see text).
Also shown is the sensitivity expected for 3 live years of the 
new telescope NT200+.}\label{fig4}
\end{figure}

Fig.~\ref{fig4} shows our upper limit on 
the all-flavor $E^{-2}$ diffuse flux (\ref{eq2})
as well as the model independent limit on the resonant $\bar{\nu}_e$ flux 
(diamond) (\ref{eq3}). Also shown are the limits obtained by AMANDA 
\cite{AMANDAHE} and MACRO \cite{MACROHE} 
and theoretical bounds and predictions for diffuse neutrino fluxes
of different origin (see \cite{HEB}).

\section{Conclusion}
The Baikal neutrino telescope NT200 is 
taking data since April 1998. 
The upper limit obtained for a diffuse 
($\nu_e+\nu_{\mu}+\nu_{\tau}$) 
flux with $E^{-2}$ shape
is $E^2 \Phi = 8.1 \times 10^{-7}$cm$^{-2}$s$^{-1}$sr$^{-1}$GeV.
The limits on fast magnetic monopoles and on  
a diffuse $\bar{\nu_e}$ flux
at the resonant  energy 6.3$\times$10$^6$GeV are presently the most
stringent ones.


 {\it This work was supported by the Russian Ministry of Education and Science, the 
  German Ministry of Education and Research and the Russian Fund of Basic Research
  (grants 05-02-17476, 05-02-16593, 07-02-10013, 07-02-00791), by the Grant of
  the President of Russia NSh-4580.2006.2 and by NATO-Grant
  NIG-9811707 (2005).}



\end{document}